\documentclass[graybox,natbib,nosecnum]{svmult}
\bibpunct{(}{)}{;}{a}{}{,} 

\pdfoutput=1   

\usepackage{mathptmx}       
\usepackage{helvet}         
\usepackage{courier}        
\usepackage{type1cm}        
\usepackage{multirow}                            
\usepackage{supertabular}                            
\usepackage{longtable}                            
\usepackage{amssymb}                            

\usepackage{makeidx}         
\usepackage{graphicx}        
\usepackage{multicol}        
\usepackage[bottom]{footmisc}
\usepackage[normalem]{ulem}	
\usepackage{hyperref}  
\usepackage{natbib}

\usepackage{soul}   

\newcommand{\RE}{\ensuremath{R_{\oplus}}}

\newcommand{\totalLC}{177\,454}
\newcommand{\totaltargets}{163\,665}
\newcommand{\totaldw}{101\,083}
\newcommand{\totalV}{61\,174}
\newcommand{\totalEBfilt}{1653}
\newcommand{\totalCB}{616}
\newcommand{\totalmono}{116}

\newcommand{\totalCandfilt}{594}
\newcommand{\totalbinaries}{2269}

\newcommand{\totalGh}{211}

\newcommand{\totalFA}{824} 

\begin{document}

\title*{CoRoT: a first space-based transiting survey to explore the close-in planets populations}
\author{Magali Deleuil and Malcolm Fridlund}
\authorrunning{The CoRoT exoplanet program} 
\institute{Magali Deleuil \at Aix Marseille Universit\'e, CNRS, CNES, LAM (Laboratoire d'Astrophysique de Marseille) UMR 7326, 13388, Marseille, France, \email{magali.deleuil@lam.fr}
\and Malcolm Fridlund \at Leiden Observatory			
P.O. Box 9513
NL-2300 RA Leiden
The Netherlands \email{fridlund@strw.leidenuniv.nl}}
\maketitle

\abstract{The {CoRoT} (COnvection, internal ROtation and Transiting planets) space mission 
was launched in the last days of 2006, becoming the first major space mission dedicated to 
the search for and study of exoplanets, as well as doing the same for asteroseismological studies 
of stars. Designed as a small mission,
it became highly successful, with, among other things discovering the first planet proved 
by the measurements of its radius and mass to be definitely ''Rocky'' or Earth like in its 
composition and the first close-in brown dwarf with a measured radius.
Designed for a lifetime of 3 years it survived in a 900~km orbit around the Earth for 6 years discovering in 
total 37 planetary systems or brown dwarfs, as well as about one hundred planet candidates 
and \totalbinaries\ eclipsing binaires, detached or in contact. 
In total CoRoT acquired 
177 454 light curves, varying in duration from about 30 - 150 days. 
CoRoT was also a pioneer in the organisation and archiving of such an exoplanetary 
survey. \\
The development and utilization of this spacecraft has left a legacy of knowledge, both as what concerns the scientific objectives as well as the technical know-how, that is currently being utilized in the construction of the European CHEOPS and PLATO missions.}

\section{Introduction }
The {CoRoT} mission is a space observatory launched by CNES, the French space agency in December 2006. It has its roots in the new science of {asteroseismology}, which in itself originated from {helioseismology}, the study of the microvariations of our Sun.  Such observations could literally Òlook inside the SunÓ and determine parameters such as density profiles, internal rotation and age among other things.

The mission was first proposed in 1993, when CNES issued a call for ideas  for what they called {\it small missions}. This gave French scientists the opportunity to propose and develop a much more ambitious mission than the smaller asteroseismology instrument EVRIS that had already flown on a Russian spacecraft. This was the origin of CoRoT, devoted to the study of stellar COnvection and internal ROTation. The original objective of CoRoT was to carry out very high-precision observations of stellar oscillation mode frequencies, for a dozen of bright solar to F-type stars, in order to detect p-modes (microvariations where pressure, p, is the restoring force) in order to obtain constraints for models of internal structure, and to begin to quantify the internal rotation of stars other than the Sun. CNES selected the project at the end of 1994 for a launch in 1998! The detection of the first exoplanet, {51 Peg}, \cite{MayorQueloz1995} led to the realization that the CoRoT requirements should also allow the detection of {transiting exoplanets}. The detection of transiting planets was added in 1997 to the new scientific program of CoRoT  whose name was changed to COnvection, internal ROtation and Transiting planets.

Different financial and administrative problems led to enlarge the participation to other countries: Austria, Belgium, Brazil, Germany, and Spain, and the {ESA} Science Program decided to contribute to the project, giving to CoRoT an European and even wider impact. The final mission selection took place only in 2000, with a launch foreseen in 2006 after a development phase that started in 2003. The latter is described fully in \cite{2016cole.book...11B} and \cite{Fridlund2006} and will not be detailed here. The development time was 4 years only, short compared to what is usually required, but the mission succeeded in being launched on December 27, 2006. CoRoT was placed very accurately (errors of order a few hundred meters) by the Soyuz/Fregat launch vehicle into the desired orbit. The launcher was the first version of the Soyuz-Fregat rocket later to be used by the ESA as its medium-size launch vehicle \citep{2006ESASP1306..255L}. The fact that the Fregat stage could steer CoRoT into the exact orbit saved enough propulsion capacity to last for the whole extended mission. The spacecraft was placed into orbit in perfect condition. The tests and evaluations of the spacecraft used up only half the allotted time and, on January 17, 2007, observations of the first field began.

\begin{figure*}[ht]
\begin{center} 

\includegraphics[scale=0.4]{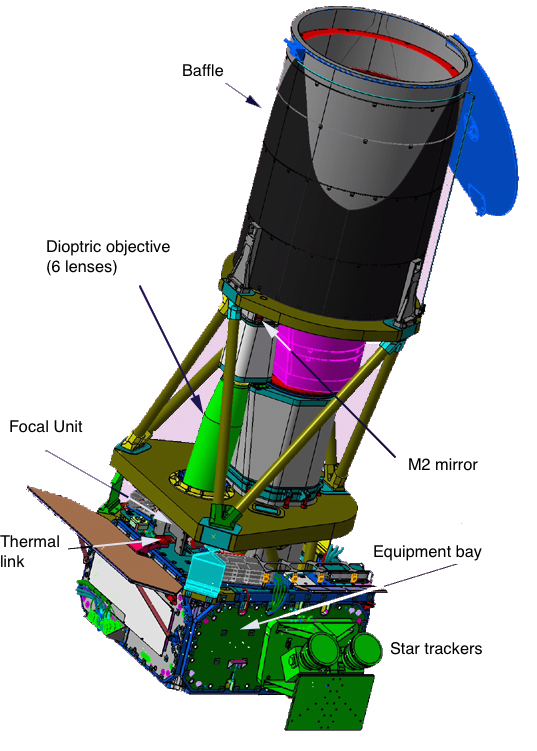}
\caption{Satellite overview. \label{spacecraft}}
\end{center}
\end{figure*}

\section{The Spacecraft}
The CoRoT  satellite was sent into a circular polar orbit with an altitude of 896~km and remained operative there until November 2, 2012, when a computer error terminated the mission.
CoRoT  was designed as a low-cost mission utilizing a proven spacecraft bus of the {PROTEUS} family (of which 5 were manufactured for a number of missions) allowing a faster and cheaper development. Together the plateform and the instrument measure 4.2~m along the longest dimension and with a launch wet mass of 626~kg, and the payload comprising 300~kg was thus relatively small. The payload consisted of a 27~cm off-axis telescope, the associated camera, and the mechanical structures and electronics (Fig~\ref{spacecraft}). The spacecraft bus consumed 300 W of power, while the payload required another 150 W \citep{2006ESASP1306..103L}.

In order to comply with its scientific objectives, the instrument had to deliver a very stable signal. This stringent stability requirement implied: (i) a high level of straylight rejection, most of it due to the nearby Earth, (ii) a high pointing stability, and (iii) a high level of performance for the thermal control subsystem. The stability requirements associated with this required the use of hyper stable materials to control the various sources of noise and ensure to reach the photon noise limit \citep{2006ESASP1306...19B}. The opto-mechanical design for the telescope, its control of jitter, and its high-performance compact baffling concept were implemented for the first time and its performance has now been fully demonstrated in flight. 

\begin{figure*}[ht]
\begin{center} 
\includegraphics[scale=0.90]{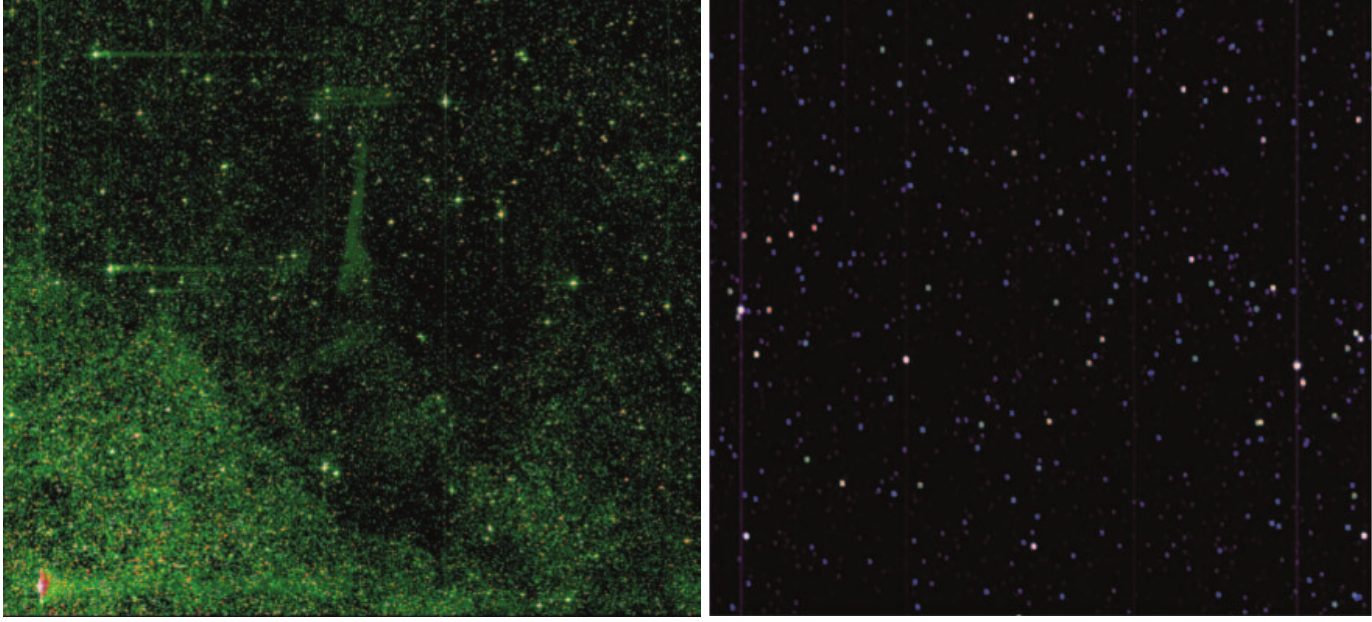}
\caption{The sky on one of the faint star channel (left)  and one of the bright star channel (right) (Chaintreuill, {\it priv. com.}). The prism in front of the CCDs confers this slightly blurred aspect to the stars of the faint channel, while those in the bright channel are simply defocused. \label{Field}}
\end{center}
\end{figure*}

To accommodate the two prime scientific objectives, instead of having two separate instruments, the adopted approach consisted in splitting the focal plane in two parts, each dedicated to one of the mission goal. At that time, the exoplanet and seismology observations aimed at targeting stars of different brightnesses. Indeed, while achieving the detection of  solar oscillations with a precision of the order of 0.1 $\mu$Hz required to match the photon noise on bright stars with a very high temporal cadence of one measurement every second, for the transit detection the low transit probability required to observe thousands of stars in order to increase the chance of detection. To fulfill this requirement with the limited field of view of the instrument, the exoplanet program targeted stars in the range 11 to nearly 16. A pair of CCDs was thus dedicated to each program and they could not be interchanged. The asteroseismology program concentrated on very bright stars, typically in the magnitude range 6 to 9. The CCDs for the asteroseismology program were defocused while the exoplanet CCDs were on focus but with a small biprism inserted above the devices (Fig.~\ref{Field}). The resulting point spread function (PSF) in the faint star channel is an on-axis spectrum at a very low spectral resolution. The goal was to provide a chromatic information in order to disentangle true planetary transits from stellar activity features, like spots, or background eclipsing binaries \citep{Rouan1999}. 

The two onboard data processing units (DPU) controlled each two CCDs at the time but one for asteroseismology and one for exoplanetology, thereby providing redundancy. The breakdown of the first data processing unit (DPU1), which occurred in March 2009, caused the loss of one CCD in each of the exoplanet and seismology channels and reduced the field-of-view by half but didnÕt cause the stop of one of the programs.

In the exoplanet channel, the onboard processing and telemetry capacity of the satellite enabled the observation of up to 6000 target stars per CCD. At the start of each observing run, an image of the complete field of view was obtained and downloaded to the ground. Based on this image, each target star was automatically assigned a photometric aperture selected from a library of 254 predefined masks, built so as to optimize the signal-to-noise ratio of the integrated flux \citep{Llebaria2006}. 
For these target stars, the photometry was carried out onboard, and only the light curves were downloaded to Earth. In addition, twenty 10 by 15 pixels windows were downloaded from each CCD in order to provide sky reference images and monitor changes in the background level. A further 80 such windows, designated "imagettes'', initially foreseen as calibration, were assigned to selected special (bright) targets of interest and were downloaded as pixel-level data enabling a dedicated photometric analysis on the ground. The nominal magnitude range of the mission, for the targets in the exoplanet channel, were of magnitude 11 to 16, but a number of brighter stars were also observed, despite being saturated, and the data from most of these were downloaded as imagettes allowing a more precise photometry to be optimized in later processing.

For stars with magnitude 15 or brighter, the photometric aperture was divided along detector column boundaries into three regions of the PSF corresponding approximately to the red, green, and blue parts of the visible spectrum. This way three color light curves were acquired for each such object for up to 5000 stars per CCD. These color light curves were summed together on the ground to give a corresponding {\it white} light curve. For stars with magnitudes larger than 15 only, white light curves were extracted, and no color information was available. 

 The cadence of the observations could be set to either 32~sec or 512~sec in the faint channel mode, while the asteroseismology channel allowed settings down to 1~sec. The basic faint channel integration time was 32~sec, but the flux of 16 readouts was coadded on board over a 8.5~min time span before being downloaded to accommodate with the telemetry budget. The nominal sampling time of 32~ sec was however preserved for 1000 selected targets (500 per CCD), known as oversampled targets. These  targets were selected at the beginning of each run, but the list was then updated every week, thanks to a quick look analysis of the crudely processed light curves and the pre-detection of transits. 

\begin{figure*}[ht]
\begin{center} 
\includegraphics[scale=0.1]{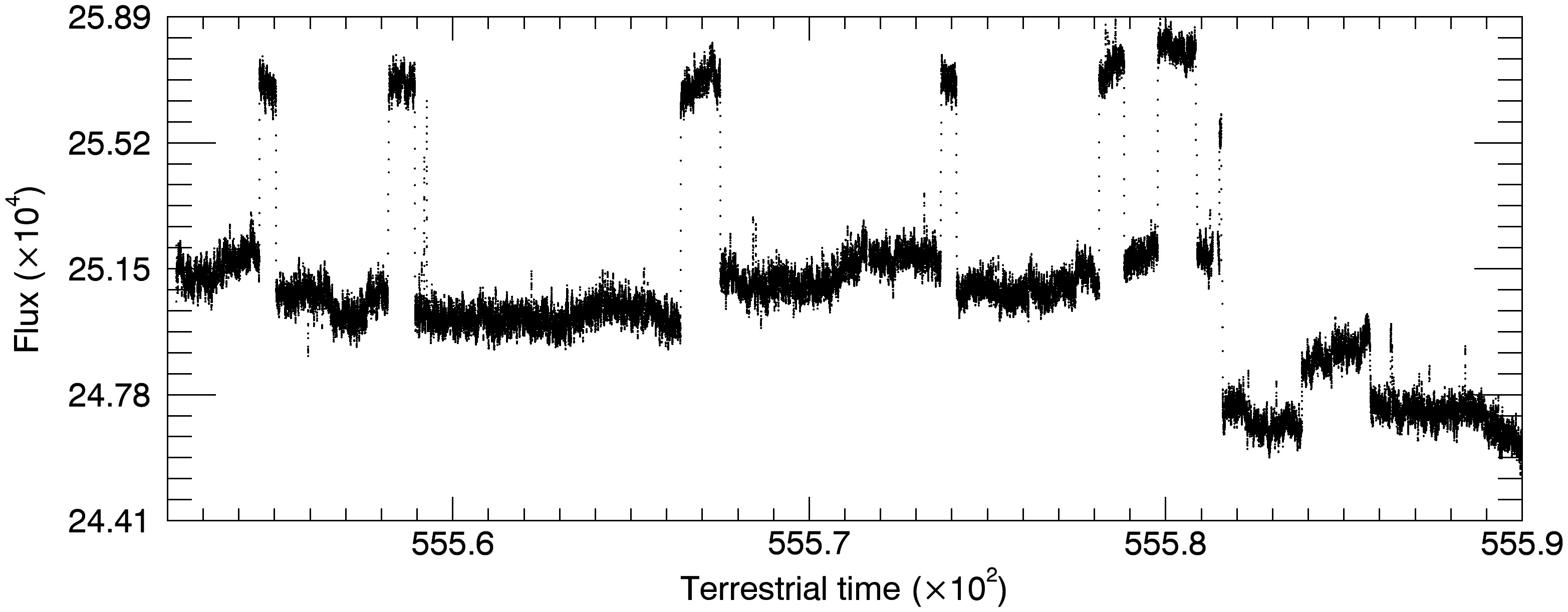}
\includegraphics[scale=0.1]{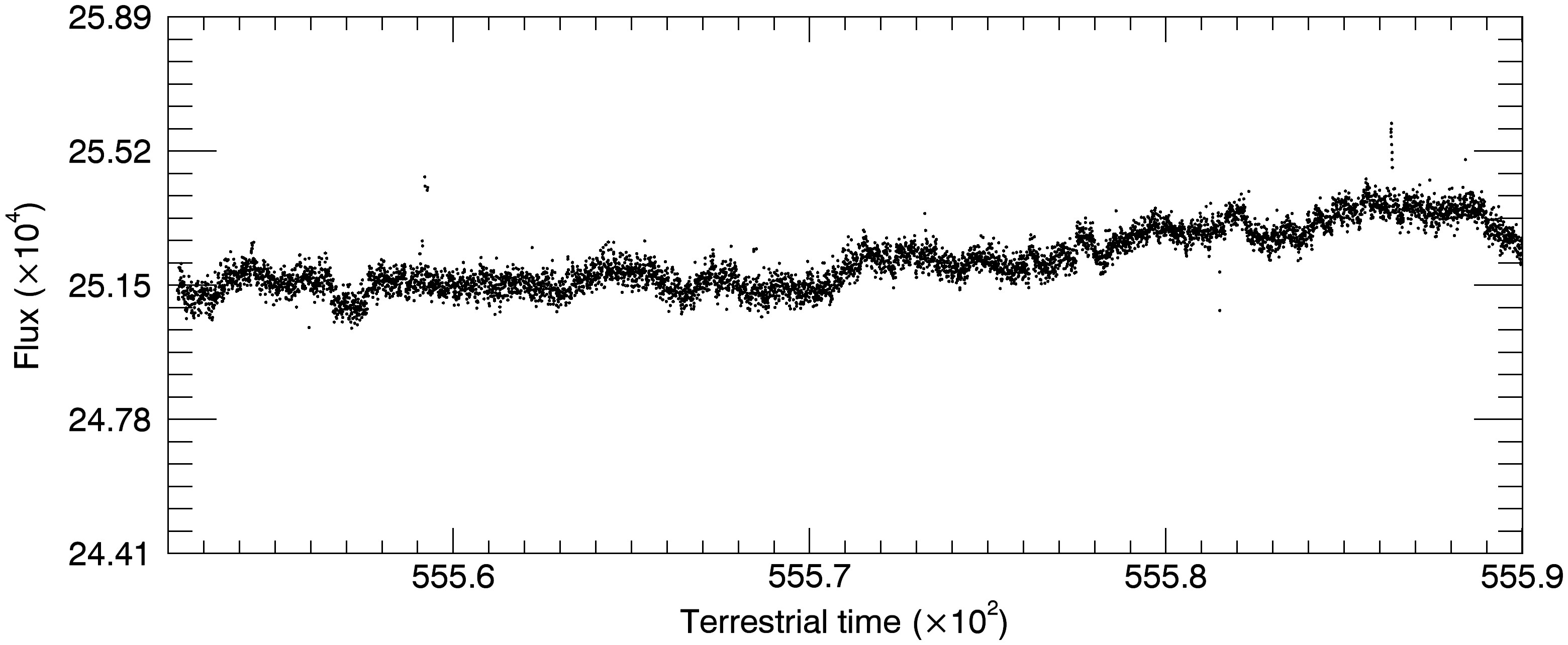}
\caption{Section of a light curve in the faint star channel before (top) and after (bottom) correction by the latest version of the data reduction pipeline, including systematics.  The jumps which are clearly visible on the raw light curve have been automatically detected and corrected. \label{LC}}
\end{center}
\end{figure*}

In 2016, the complete set of CoRoT light curves that is those from both the bright and faint channels were homogeneously processed with the latest version of the pipeline and released to the community \citep{Chaintreuil2016B}. A complete description of the different steps of the final data reduction pipeline and of the associated algorithms is provided in \citep{Ollivier2016}. In addition to the regular corrections, such as crosstalk or background contribution corrections, which were already included in the pipeline \citep{Auvergne2009}, but updated in this last version, new corrections were implemented. With this latest release, the user can thus get ready-to-use light curves corrected from the jumps in the photometry such as those induced by a change in temperature or by impacts of protons onto the CCD but also from systematics \citep{Guterman2016} (Fig.~\ref{LC}).
 
As mentioned above, CoRoT was based on the multipurpose PROTEUS spacecraft bus. This configuration, while saving cost and time in development, restricted the mission to a {low Earth orbit (LEO)}. Because of this limitation, and in order to be able to observe the same field of view for as long time as possible, at least up to 6 months, and at the same time without allowing either too much scattered solar light to enter the telescope, or experience too many occultations by the Earth, the satellite had to be injected into a polar orbit and restricted to observe along line of sights roughly perpendicular to the orbital plane. Every six months (in April and October), to avoid  blinding by the Sun, the satellite was rotated by 180~deg with respect to the polar axis and a new observation period started in the opposite direction. As a consequence, the continuous viewing zones of CoRoT were two almost circular regions of 10$^o$ radius, called the {CoRoT eyes}. They are centered on the galactic plane at 6h 50d (near the galactic anti-center) and 18h 50d (near the galactic center), respectively. 

\section{The Scientific Organisation}

In CoRoT, two scientific topics were coexisting in the same instrument. The asteroseismology segment had as its objective to divulge the internal physical parameters of stars for the first time. The objective of the exoplanet element was to discover new transiting exoplanets, to measure their diameter with unprecedented precision and further determine their properties. 
While these two objectives  could at first appear very different, it is actually true that both the technology behind the actual measurements, namely ultrahigh-precision photometry, as well as the science in them, have a deep connection. One of the superb achievements brought by the CoRoT results is that the understanding of exoplanets is based on an equally deep understanding of the host star. Planets are orbiting a host star, the physics of which governs their evolution and properties. But the opposite is also true. During the formation phase of a planetary system both star and planets are tightly connected through the transfer of angular momentum and through chemical changes in the accretion disk that must take place as star and planets accrete from the original material.

Nevertheless, we will here mainly discuss the exoplanetary part of the mission.

The CoRoT project was committed to deliver to the scientific community light curves that were properly reduced and corrected for the main instrumental defects and ready for scientific analyses. The detection of planetary transits was thus left to the discretion of the science team with the exception of a real-time detection carried out in the {Alarm Mode} on a weekly basis \citep{Quentin2006,Surace2008}. The Alarm Mode was carried out to retune the cadence of the observations of that particular star to sample the light curve every 32s instead of 512s, allowing the possibility to observe forthcoming transit events with higher temporal resolution. In order to interpret the actual transit shape in terms of physical parameters of both the host star and the planet it is imperative to have the highest temporal resolution. In addition, this real-time detection allows  to save time in the follow-up process of the planet candidates.

During the years that preceded the launch of the satellite, the scientists who were involved in the exoplanetary program of the mission made the decision to work as a single international team. The goal was to share the workload and results so that to increase the scientific return of the mission and to avoid time wasted in competition. This team thus came to consist of individual scientists from all the nations who were partners in the project and including members from ESA\,s science department. It took the name of {CoRoT Exoplanet Science Team (CEST)} and organized every aspects of the analysis, starting from the transit-like features detection to the detailed analysis of the planet's properties. One important aspect of this collaborative work was the follow-up observations of planet candidate\,s.  The CoRoT exoplanet program has been indeed supported by a large accompanying ground-based observation program \citep{Deleuil2006}. Operating more than a dozen of telescopes in various places, Europe, Hawaii, Israel, and Chile, with size varying from 1~m to 8~m, the team used various techniques: photometric observations \citep{Deeg2009}, high-contrast imaging \citep{Guenther2013}, and spectroscopy including radial velocity measurements \citep{Bouchy2009}. The goal was to identify false positives, to fully secure planets, and to determine their complete set of parameters in order to derive the planetary properties.
\begin{table}
\caption{Overview of the CoRoT observed fields and targets. Column 5 gives the number of stars monitored during the pointing, column 6 the number of those targets that were observed in this field only, column 7 the number of those targets that were observed in another field, column 8 the number of those targets that were observed in two others fields, column 9 the number of targets classified as dwarfs.}
\label{tab:1}      
\begin{tabular}{lcrcrrrrr}
\hline\noalign{\smallskip}
Field  &   CCD  & Duration &  Overlap &  Observed & Targ. $\sharp$ 1 & Targ. $\sharp$ 2 & Targ. $\sharp$ 3 &  Dwarfs \\
          &                 &  (days)   &                 &        targets     &                             &                            &                           &  (IV/V)  \\
\noalign{\smallskip}\svhline\noalign{\smallskip}
IRa01 & 2 & 54.3 & LRa01/LRa06 & 9921 & 8216 & 821 & 884 & 6550 \\
LRa01 & 2 & 131.5 & IRa01/LRa06 & 11448 & 11448 & 0 & 0 & 8961 \\
SRa01 & 2 & 23.4 & SRa05 & 8190 & 5822 & 2368 & 0 & 4218 \\
SRa02 & 2 & 31.8 & LRa07 & 10305 & 10305 & 0 & 0 & 7990 \\
LRa02 & 2 & 114.7 &               & 11448 & 11448 & 0 & 0 & 9410 \\
LRa03 & 1 & 148.3 &               & 5329 & 5329 & 0 & 0 & 3862 \\
SRa03 & 1 & 24.3 &               & 4169 & 4169 & 0 & 0 & 3038 \\
LRa04 & 1 & 77.6 &               & 4262 & 4262 & 0 & 0 & 2967 \\
LRa05 & 1 & 90.5 &               & 4648 & 4648 & 0 & 0 & 3332 \\
SRa04 & 1 & 52.3 &               & 5588 & 5588 & 0 & 0 & 3840 \\
SRa05 & 1 & 38.7 & SRa01 & 4213 & 4213 & 0 & 0 & 2452 \\
LRa06 & 1 & 76.7 & LRa01/IRa01 & 5724 & 1356 & 3484 & 884 & 947 \\
LRa07 & 1 & 29.3 & SRa02 & 4844 & 4390 & 454 & 0 & 3173 \\
SRc01 & 2 & 25.6 &               & 7015 & 7015 & 0 & 0 & 4484 \\
LRc01 & 2 & 142.1 &               & 11448 & 11448 & 0 & 0 & 4922 \\
LRc02 & 2 & 145 & LRc06/LRc05 & 11448 & 11448 & 0 & 0 & 6239 \\
SRc02 & 2 & 20.9 &               & 11448 & 11448 & 0 & 0 & 3477 \\
LRc03 & 1 & 89.2 &               & 5724 & 5724 & 0 & 0 & 3639 \\
LRc04 & 1 & 84.2 & LRc10 & 5724 & 5724 & 0 & 0 & 4200 \\
LRc05 & 1 & 87.3 & LRc06 & 5724 & 5724 & 0 & 0 & 2456 \\
LRc06 & 1 & 77.4 & LRc02/LRc05 & 5724 & 3836 & 1880 & 8 & 2029 \\
LRc07 & 1 & 81.3 & LRc08/LRc10 & 5724 & 3953 & 1771 & 0 & 1784 \\
SRc03 & 1 & 20.9 & LRc02/LRc06 & 652 & 85 & 559 & 8 & 0 \\
LRc08 & 1 & 83.6 & LRc07/LRc10 & 5724 & 5724 & 0 & 0 & 2658 \\
LRc09 & 1 & 83.6 &               & 5724 & 5724 & 0 & 0 & 2630 \\
LRc10 & 1 & 83.5 & LRc04/LRc07 & 5286 & 4618 & 668 & 0 & 1825 \\
\hline
Total &  &  &  & 177454 & 163665 & 12005 & 892 & 101083 \\
\noalign{\smallskip}\svhline\noalign{\smallskip}
\end{tabular}
\end{table}

\section{Exoplanetary science with CoRoT}

In total, CoRoT made observations of 163 665 targets over 26 stellar fields in the faint star channel. The various fields are identified by five digits; the two first refer to the duration of the run but the first run named as Initial Run (IR): LR for runs with a duration longer than 70 days (Long Runs) and SR for runs with a much shorter duration (Short Run). The third digit "a" or "c", means galactic anticenter or center direction respectively. Finally, the last two digits are just the chronological order of the given type of field in a given direction.

According to the revised exoplanet input catalog \citep{Damiani2016}, \totalV\ of these targets have been classified as luminosity class V. In the CoRoT exoplanet context, we were interested in Òdwarf starsÓ, i.e., luminosity classes IV and V, and then a total of \totaldw\ stars were available to the mission. The dwarf/giant identification was based on a simple color-magnitude diagram \citep{Deleuil2009, Damiani2016}, and while this is reliable from a statistical point of view, individual targets could be misclassified. The scope of the preparatory classification and of the targets selection was thus to ensure the observation of all possible dwarfs, at least those with a size suitable for planetary transit detection in a given pointing. Nevertheless, in a given field, the number of stars of classes V and IV could not populate all the available photometric windows, and thus the remaining ones were allocated to targets selected for complementary science programs.  There are however noticeable differences in the ratio of dwarf stars to giants from one field to another due to variations in the stellar populations, given their position in the Galaxy and from different reddening between the fields (Table~\ref{tab:1}). 
\begin{figure*}[ht]
\begin{center} 
\includegraphics[scale=0.5]{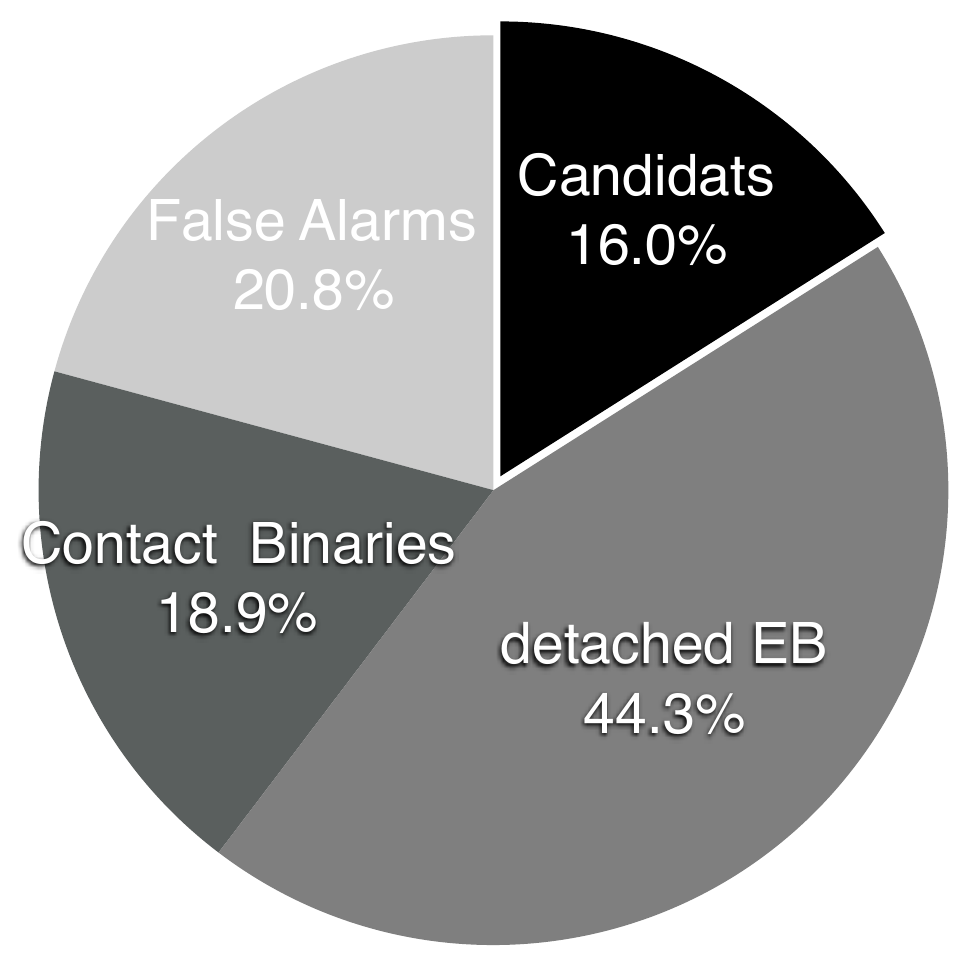}
\caption{Distribution of the detected transit-like events among planet candidates (Cand), detached eclipsing binaries (EB), false alarms (FA), and contact binaries (CB). \label{Distribdets}}
\end{center}
\end{figure*}

CoRoT observations resulted in \totalLC\ light curves whose duration ranges from 5.0 days for the stars observed in the SRc03 up to 148.3 days for those that were observed in the LRa03. The SRc03 field was a special pointing dedicated to the sole observation of one further transit of {CoRoT-9b} \citep{Deeg2010,Lecavelier2017,Bonomo2017}. Not only was its duration very short, but the number of targets was also very limited (652 only) all of which had been targeted by a previous observation. Table~\ref{tab:1} provides a summary of the CoRoT runs and targets that were observed in each of them.  Among the \totaltargets\ targets, 12 005 were observed twice and 892 three times. 

More than 4000 transit-like features were detected in total by the detection teams. Their depth could be as low as 0.01\% and periods range from 0.27 to 95 days. Among the detected transit-like features, \totalmono\ displayed only 1 transit in any given run. Their duration ranges from nearly 2 hours up to 112.16 hours for the longest event. From this list, subsequent analyses identified a total of \totalFA\ false alarms of various kinds. This category includes simple false alarms due to errors in the detection software or to a discontinuity in the light curve (see Fig.~\ref{LC}), but also at least \totalGh\ signals due to a bright eclipsing binary whose light leaked over one or more pixel columns and left its photometric imprint in the light curve(s) of other nearby target(s). Further \totalbinaries\ eclipsing binaries among which \totalCB\ are contact and \totalEBfilt\ are detached binaries, as well as a total of about 600 transit events initially classified as planet candidates were detected \citep{Deleuil2018}. Figure~\ref{Distribdets} shows the distribution of the detected transit-like features over the different classes of events. 

\begin{figure*}[ht]
\begin{center} 
\includegraphics[scale=0.5]{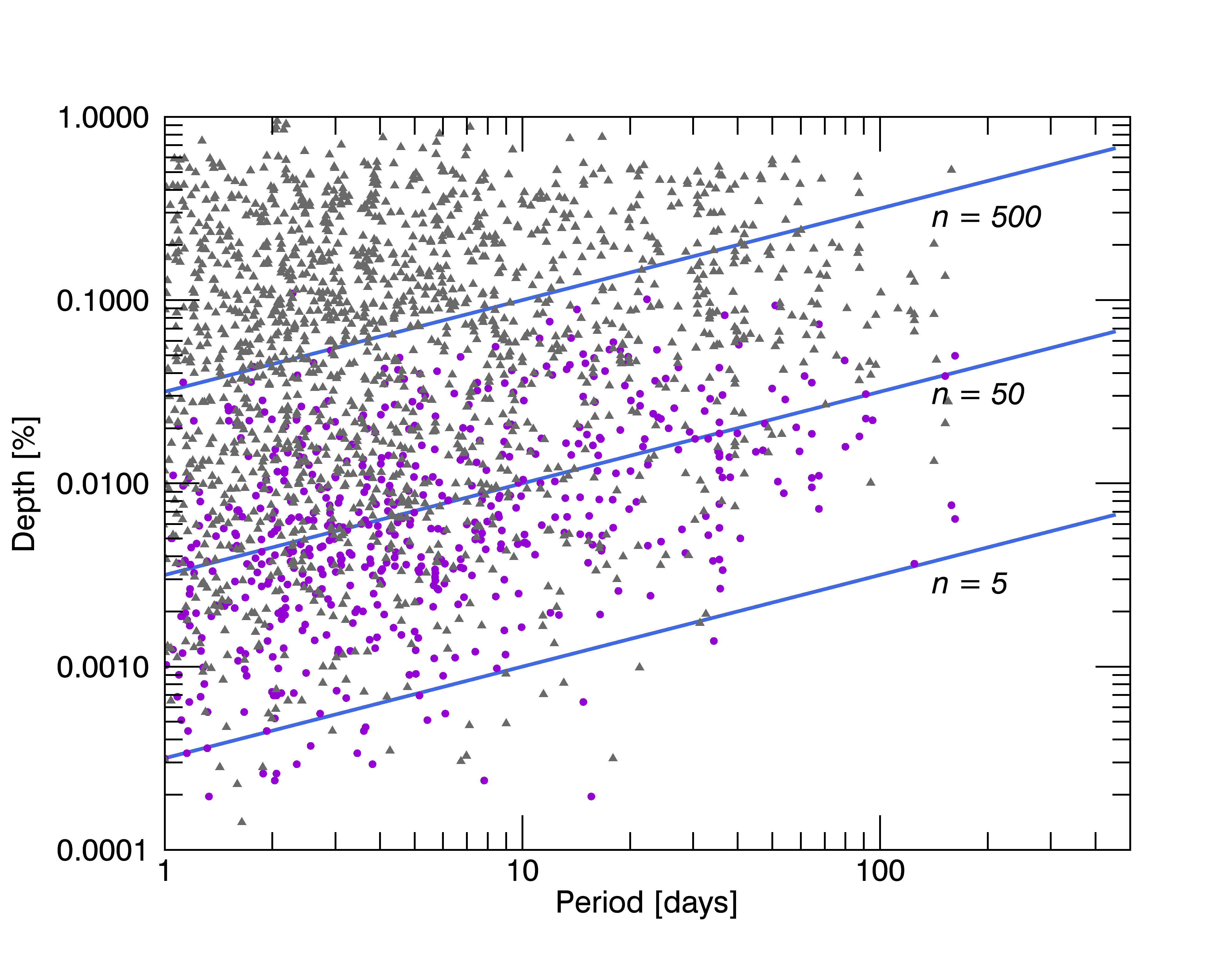}
\caption{Period-depth diagram of all candidates (violet circle) and EB (gray triangle) detected in the CoRoT fields. The plain lines show the transit signal to noise ratio (SNR).\label{DepthPeriod}}
\end{center}
\end{figure*}

Out of the planet candidates, 37 exoplanets and brown dwarf systems have been confirmed, with 2 multiplanet systems only. These numbers appear paltry when comparing with the thousands of exoplanets that have been confirmed subsequently, particularly through the {Kepler mission} and the succeeding {K2} mission. The reason for these differences is mostly due to the difference in sensitivity between the two instruments (see Fig.~\ref{DepthPeriod}).

\begin{figure*}[ht]
\begin{center} 
\includegraphics[scale=0.05]{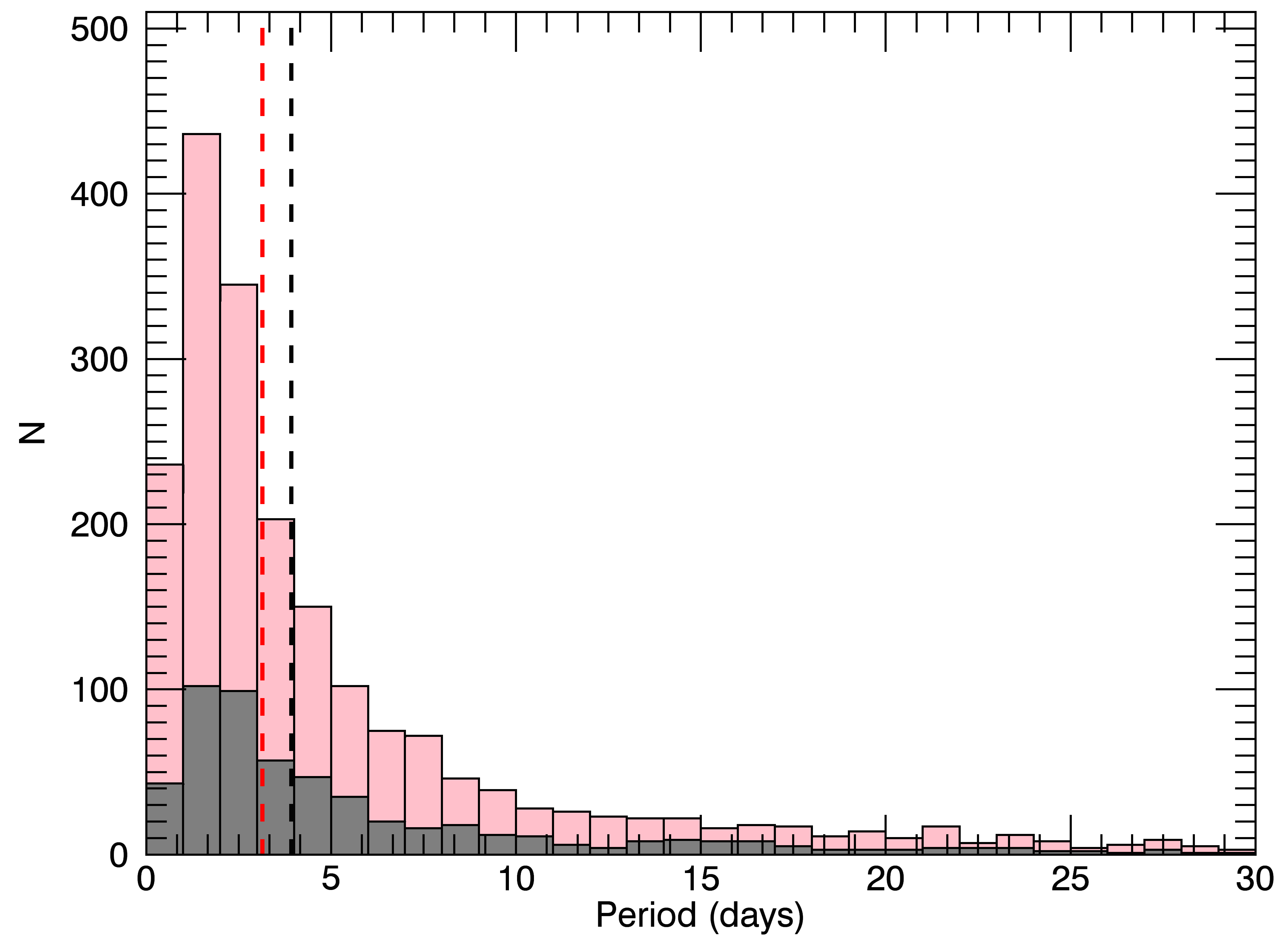}
\caption{Period distribution (stacked histograms) of EBs (pink) and candidates (gray). The dash lines give the median of each distribution. \label{fig:periods}}
\end{center}
\end{figure*}

With typical durations of the runs of 69.4 days for the fields in the anticenter and 83.6 days for those in the center, CoRoT has been well adapted for the exploration of the close-in giant population. Two thirds of both the candidates and the EBs have orbital periods shorter than 10 days with a peak value of 1.5 day, and 90\% shorter than 25 days (Figure~\ref{fig:periods}). Transit events at orbital periods in excess of 100 days were also reported. Some of those are single transit events, but some were detected as periodic events during long runs, as was the case for CoRoT-9b \citep{Deeg2010}.

The CoRoT team followed the principle of obtaining both a firm mass and a precise radius, the latter owing to the exquisite photometry delivered by the instrument and the former to the great amount of follow-up work carried out. The ensemble of CoRoT planets is thus a very good sample of bodies on which to test theories of planetary structure as well as planetary formation theories.

\section{Detection and vetting}
While at the time of writing (December 2017) there are some robust planet candidates still in the final stage of the validation process, as illustrated by the recent publication of {CoRoT-32b} \citep{Boufleur2018}, CoRoT currently accounts for 38 transiting planets detected and securely confirmed. These are all fully characterized thanks to comprehensive and systematic ground-based follow-up program.  Of these 38 objects labeled as planets, there are in fact 2 brown dwarfs, {CoRoT-15b} \citep{Bouchy2011} and {CoRoT-33b} \citep{Csizmadia2015} and one object, {CoRoT-3b} \citep{Deleuil2008}, whose exact nature, light brown dwarf or massive planet, depends on how these objects are defined, something that remains the subject of controversies \citep[see][]{Schneider2011, Hatzes2015} and the Chapter "Definition of Exoplanets and Brown Dwarfs'').

Follow-up observations significantly increased the scientific return of the mission but required a large effort on various facilities. It involved ground-based photometry taken during and just outside the transits with larger telescopes at higher spatial resolution and spectroscopic observations of the host star. The former was used to verify whether the transits occurred on the main target or a fainter nearby star. High-contrast imaging helped also identify background binaries or physical triple systems further. Radial velocity measurements enabled identifying objects with multiple sets of spectral lines and to measure the masses of any actual planets, together with the eccentricity of their orbits. Spectra collected for RV measurements or additional spectroscopic data were also used to estimate the host star\,s parameters: effective temperature, gravity, metallicity, and further deduce their mass, radius, and age.
\begin{figure*}[ht]
\begin{center} 
\includegraphics[scale=0.5]{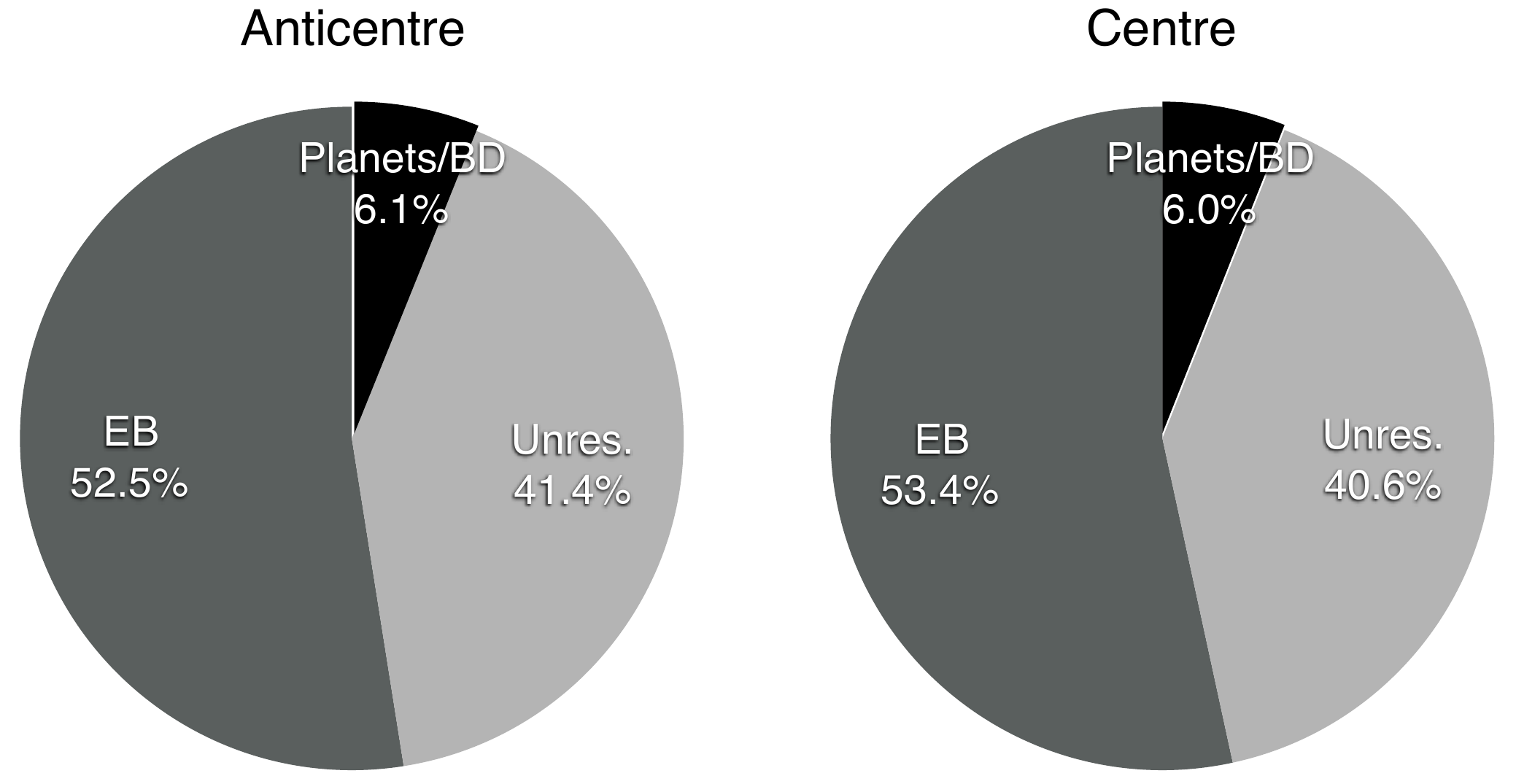}
\caption{Distribution of all features initially selected as candidates among: confirmed planets and brown dwarfs (P/BD), detached eclipsing binaries (EB) finally identified through follow-up observations or a deeper analysis of their light curve, and candidates whose nature still remains unknown (Unres.) in the two pointing directions of CoRoT. \label{resolv}
}
\end{center}
\end{figure*}
Follow-up observations started as soon as possible after the end of each run and even while CoRoT observations were still ongoing thanks to detections done by the Alarm's software.  Among the \totalCandfilt\ candidates which were at one point deemed worthy of follow-up,  a bit more of 70\% were observed by at least one ground-based facility. Because follow-up observations were triggered as soon as possible while the candidates vetting and light curves modeling continued for months with a process which evolved continuously during the mission, 88 of the candidates which were initially included in the follow-up program were later discarded on the basis of a refined light curve analysis. In some cases, for example, secondary eclipses that indicate an EB were later detected in the full CoRoT light curve of what first appeared as an interesting candidate. Meanwhile the candidate had however already been observed. Follow-up observations concentrated on the brightest stars, with  a completeness close to 100\% for candidates with r-mag $<$ 14. Much fainter targets were also followed-up, but the fraction drops to 78\% for 14 $\le$ r-mag $<$ 15 and 63\% for the faintest targets with r -mag $>$~15. 

Figure~\ref{resolv} shows how the transit features initially classified as candidates finally distribute after follow-up observations or a deeper analysis of their light curve. The {\sl eclipsing binary} class gathers the cases where the source of the transiting signal coincides with the target, which includes spectroscopic binaries and configurations in which the source of the transit has been identified as an eclipsing binary other than the target, but whose light contributes to the target measured in CoRoT\,s photometric aperture. These we denoted also as CEB for {contaminating eclipsing binaries (CEB)}. The {\sl unresolved} category comprises candidates that were either not followed-up or those whose follow-up observations remained inconclusive. The later correspond to cases where the host star is either a hot star or a fast rotator, which prevents assessing the nature of the detected companion with the usual current techniques, but also some  bright targets for which ground-based photometry didn't point out any contaminating star as the source of the signal but for which repeated RV observations reveal no significant variation consistent with the ephemeris of the transits. Confirmed planets/brown dwarfs account for only 6\% of the initial list of candidates. 

Considering only the candidates whose nature has been resolved, from both follow-up and light curves detailed analysis, on-target EB account for 53.7\% of them, CEB for 36.1\%, and planets/brown dwarfs are 10.2\%. This gives a false positive rate close to 90\% for CoRoT over the complete mission.

Among the candidates that remain unresolved, a bit more of 100 still present some chances of being of planetary nature.  Nearly half of them were subject to ground-based complementary observations which however remained inconclusive about their exact nature or would just require some deeper analysis as exemplified by CoRoT-32b. For the other half, the faintness of the targets (usually r-mag $>$ 14) did not allow a proper characterization of the transiting body. Indeed, even for Jupiter-size planets, radial velocity measurements remain difficult for targets with magnitudes greater than 14.5.  Taking into account the spectral classification of these targets, the results achieved for candidates, and assuming that the unresolved cases follow the same distribution as those whose nature has been secured, one can estimate that 8 $\pm$ 3 planets are still to be identified as such. This highlights the challenge of establishing the planetary nature of faint candidates which are still challenging the current spectrograph performances.This experience also motivates the search for transiting planets on brighter stars, as will be done by the TESS and PLATO missions.

\section{The exoplanetary Yield - Giants and the first Rocky Planet}

With its high-precision photometry and its typical pointing duration of a few tens of days, CoRoT was well adapted to explore the close-in planet population, and its contribution to the properties of this population has been pioneering.  
Table~\ref{tab:all} provides a summary of parameters of the published CoRoT planetary systems. 
While much more modest than Kepler, CoRoT accounts for number of first results on the properties of close-in transiting populations, a large fraction of them being highlighted in \cite{Moutou2013}: 
\begin{itemize} 
\item[-] The first terrestrial planet, {CoRoT-7b}  \citep{Leger2009, Queloz2009}.
\item[-] The first transiting substellar companion/brown dwarfs, CoRoT-3b \citep{Deleuil2008} and CoRoT-15b \citep{Bouchy2011} well in the gap between planets and low mass stellar companions. 
\item[-] The first temperate Jupiter-like planet, CoRoT-9b \citep{Deeg2010}
\item[-] The first massive planet around a fast {CoRoT-11} \cite{Gandolfi2010}
\item[-] The first phase curve and planet occultation, {CoRoT-2b} \citep{Snellen2009, Alonso2009} 
\item[-] The first mapping of active regions at the stellar surface, {CoRoT-4b} \citep{Lanza2009B} 
\item[-] the first detection of the ellipsoidal and the relativistic beaming effects for substellar companion, CoRoT-3b \citep{Mazeh2010}
\end{itemize}
In total, CoRoT accounts today for 39 planets, with the last 4 still to be published (Grziwas et al., {\sl in prep.}). Among them, 33 are Saturn- and Jupiter-like planets. For these massive planets, CoRoT light curves have enabled detailed analyses on their properties. It allowed to probe their interior composition and to quantify their various enrichment in heavy elements  \cite[][e.g.]{Borde2010,Deleuil2012}. Close-in planets with non-zero eccentricity  like {CoRoT-20b} \citep{Deleuil2012} and {CoRoT-23b} \citep{Rouan2012} bring observational constraints to ongoing tidal dissipation in planets and their circularization time \citep{FerrazMello2016}.\\
\\

\tablefirsthead{ 
  \hline\noalign{\smallskip}
  Planet & Period & R$_p$ & M$_p$ &${\it e}$&  ${\it a}$  & M$_\star$ & R$_\star$ &T$_{eff}$ & v sin i & Fe/H & P$_{rot}$ & Age\\
   & (days)& (R$_{jup})$& (M$_{jup})$& &(AU) & (M$_\odot$) & (R$_\odot$ )& (K) & (km/s) &  & (days)& (Gyr) \\
  \noalign{\smallskip}\svhline\noalign{\smallskip} }
\tablehead{\multicolumn{13}{l}{\small\sl continued from previous page} \\
   \hline\noalign{\smallskip}
   Planet & Period & R$_p$ & M$_p$ &${\it e}$&  ${\it a}$  &M$_\star$ & R$_\star$ &T$_{eff}$ & v sin i & Fe/H & P$_{rot}$ & Age\\
   & (days)& (R$_{jup})$& (M$_{jup})$& &(AU) & (M$_\odot$) & (R$_\odot$ )& (K) & (km/s) &  & (days)& (Gyr) \\
   \noalign{\smallskip}\svhline\noalign{\smallskip} 
}
\tabletail{\multicolumn{13}{r}{\small\sl \ldots}\\\hline}
\tablelasttail{\hline}
\topcaption{The 35 confirmed and published CoRoT planets and brown dwarfs stars. The second row in each planetary entry give the accuracy of the result. Note that when asymmetrical error bars were provided, we give here the largest value}
\label{tab:all}       
\par
\begin{supertabular}{lccrrrrrrrrrr}
1b$^{1}$ & 1.5089557  & 1.49 & 1.03& 0.006& 0.0254 &0.95 &1.11&5950 &5.2 &-0.3 &10.7& \\
& 0.00000064 &0.08 &0.12 &0.012 &0.0004 &0.15 &0.05 &150 &1.0 &0.25 &2.2&\\
2b$^{2}$ & 1.7429964 & 1.47 & 3.31&   0.036& 0.0281&0.97 & 0.90 & 5625 &11.85 & -0.04 & 4.52 & 0.03-0.3\\
& 0.0000017 & 0.03 & 0.16 & 0.033 & 0.0009 & 0.06 & 0.02 & 120 & 0.5 & 0.1 & 0.14 &\\
3b$^{3}$ & 4.2567994  & 1.01 & 21.77&0.012 &0.057& 1.37 & 1.56 & 6740 & 18 & -0.02 & 4.6 & 1.6-2.8\\
&0.000004 &  0.07 & 1.0 & 0.01 & 0.003 & 0.09 & 0.09 & 140 & 3.0 & 0.06 & 0.4 &\\
4b$^{4}$ & 9.20205  & 1.19 & 0.72 & 0.27&0.090& 1.16 & 1.17 & 6190 & 6.4 & +0.05 & 8.9 & 0.7-2.0\\
& 0.00037 & 0.06 & 0.08 & 0.15 & 0.001 & 0.03 & 0.03 & 60 & 1.0 & 0.07 &1.1\\
5b$^{5}$ & 4.037896 & 1.33 & 0.47 & 0.086 & 0.0495 & 1.00 & 1.19 & 6100 & 1.0 &-0.25 &50.0 &5.5-8.3\\
& 0.000002 & 0.05 & 0.05 &0.07 &0.0003 &0.02 &0.04 &65 &1.0 &0.06 &10\\
6b$^{6}$ &8.886593 &1.17 &2.96 &0.18 &0.0855 &1.05 &1.025 &6090 &7.5& -0.20 &6.4 &2.0-4.0\\
&0.000004 &0.04 &0.34 &0.12 &0.0015 &0.05 &0.03 &50 &1.0 &0.1 &0.5\\
7b$^{7}$& 0.85359163 &0.141 & 0.017 &0.137 & 0.017         & 0.91 &0.82  &5275 &1.5 & +0.12 &23.6 &1.32 \\
              & 5.8E-7  &0.009 &0.003 & 0.094 &1.6E-4& 0.02 & 0.02 & 60    &1.0 &0.06 &0.1&0.75\\ 
7c$^{7}$&3.698&- &0.026&0&0.046&0.91 &0.82 &5275 &1.5 & +0.12 &23.6 &1.32 \\
&0.003&-&0.003&fixed&-&0.03 &0.04 &60 &1.0 &0.06 &0.1&0.75\\
8b$^{8}$ &6.21229 &0.57 &0.22 &0 &0.063 &0.88 &0.77 &5080 &2.0 &0.3 &20.0 &0.5-3.0\\
&0.00003 &0.02 &0.03 &fixed &0.001 &0.04 &0.02 &80 &1.0 &0.1 &5\\
9b$^{9}$ &95.273804 &0.94 &0.84 &0.11 &0.407 &0.99 &0.94 &5625 & $\le$ 3.5 & -0.01 &$\ge$ 14.0 &0.5-8.0 \\
&0.0014 &0.04 &0.07 &0.039 &0.005 &0.04 &0.04 &80 &1.0 &0.06 &5\\
10b$^{10}$ &13.2406 &0.97 &2.75 &0.53 &0.1055 &0.89 &0.79 &5075 &2.0 & +0.26 &2.0 &0.5-3.0 \\
&0.0002 &0.07 &0.16 &0.04 &0.0021 &0.05 &0.05 &75 &0.5 &0.07 &0.5\\
11b$^{11}$ &2.99433 &1.43 &2.33 &0.35 &0.0436 &1.27 &1.37 &6440 &40.0 &-0.03 &1.7 &1-3  \\
&0.000011 &0.03 &0.34 &0.03 &0.005 &0.05 &0.03 &120 &5.0 &0.08 &0.2\\
12b$^{12}$ &2.828042 &1.44 &0.92 &0.07 &0.0402 &1.08 &1.1 &5675 &1.0 &+0.16 &68.0 &3.2-9.4\\
&0.000013 &0.13 &0.07 &0.06 &0.0009 &0.08 &0.1 &80 &1.0 &0.1 &10\\
13b$^{13}$ &4.03519 &0.89 &1.31 &0. &0.051 &1.09 &1.01 &5945 &4.0 &+0.01 &13.0 & 0.1-3.2\\
&0.00003 &0.01 &0.07 &fixed &0.0031 &0.02 &0.03 &90 &1.0 &0.07 &5\\
14b$^{14}$ &1.51214 &1.09 &7.6 &0. &0.027 &1.13 &1.21 &6035 &9.0 &+0.05 &5.7 &0.4-8.0\\
&0.00013 &0.07 &0.6 &fixed &0.002 &0.09 &0.08 &100 &0.5 &0. &15 &\\
15b$^{15}$ &3.06036& 1.12& 63.3 &0 &0.045 &1.32 &1.46 &6350 &19 &+0.1 &3.0 &1.1-3.4\\
&0.00003 &0.30 & 4.1& fixed &0.014& 0.12& 0.31 & 200& 1.0& 0.2& 0.1\\
16b$^{16}$ &5.35227 &1.17 &0.54 &0.33 &0.0618 &1.1 &1.19 &5650 &0.5 &+0.19 &60.0 &3.7-9.7\\
& 0.0000 &0.15 &0.09 &0.10 &0.0015 &0.08 &0.14 &10 &1.0 &0.06 &10\\
17b$^{17}$ &3.7681 &1.02 &2.43 &0. &0.0461 &1.04 &1.59 &5740 &4.5 &+0.00 &20.0 &9.7-11\\
 & 0.0000 &0.07 &0.30 &fixed &0.0008 &0.1 &0.07 &80 &0.5 &0.1 &5\\
18b$^{18}$ &1.900069 &1.31 &3.47 &0.04 &0.0295 &0.95 &1.00 &5440 &8.0 &-0.10 &5.4 &0.05-1\\
&0.0000 &0.18 &0.38 &0.04 &0.0016 &0.15 &0.13 &100 &1.0 &0.1 &0.4\\
19b$^{19}$ &3.89713 &1.29 &1.11 &0.047 &00518 &1.21 &1.65 &6090 &6.0 &-0.02 &15.0 &4.0-6.0\\
&0.0000 &0.03 & 0.06 &0.045 &0.0008 &0.05 &0.04 &70 &1.0 &0.1 &5\\
20b$^{20}$ &9.24285 &0.84 &4.24 &0.562 &0.0902 &1.14 &1.02 &5880 &4.5 &+0.14 &11.5 &0.06-0.9\\
& 0.0000 &0.04 &0.23 &0.013 &0.0021 &0.08 &0.05 &90 &1.0 &0.12 &3\\
21b$^{21}$ &2.72474 &1.30 &2.26 &0 &0.0417 &1.29 &1.95 &6200 &11.0 &+0.00 &10.0 &3.0-5.0\\
& 0.0000 &0.14 &0.31 &fixed &0.0011 &0.09 &0.21 &100 &1.0& 0.1& 3.0 & \\
22b$^{22}$&9.75598&0.435&0.038&0.077&0.0920&1.099&1.136&5939&4.0&+0.17& $\simeq$ 16.0 & 3.3 \\
&0.00011&0.035&0.044&0.042&0.0014&0.049&0.09&120&1.5&0.12& & 2.0\\
23b$^{23}$ &3.6313 &1.05 &2.8 &0.16 &0.048 &1.14 &1.61 &5900 &9.0 &+0.05 &9.2 &6.2-7.7\\
& 0.0001 &0.13 &0.3 &0.02 &0.004 &0.08 &0.18 &100 &1.0 &0.1 &1.5\\
24b$^{24}$  & 5.1134. & 0.33 &  0.018& 0.  & 0.056 &  0.91& 0.86& 4950.& 2.0 &  +0.3 & $\simeq$ 29  & $\simeq$ 11 \\
                    & 0.0006  &  0.04&           & fixed  &0.002&  0.09 & 0.09 &150&1.5&0.15 & & \\  
24c$^{24}$  & 11.759 & 0.44 &0.088& 0 & 0.098&	 0.91 & 0.86 &4950 & 2.0   &  +0.3&\\
&0.0063 &  0.04&   0.035    & & 0.003& 0.09 & 0.09 &150&1.5&0.15&\\ 
25b$^{25}$&4.86069& 1.08& 0.27& 0.0 &0.0578 &1.09&1.19&6040&4.3&-0.01& 14.1 &4.5\\
&0.00006&0.1& 0.04& fixed & 0.002& 0.11&0.14&90&0.5&0.13& 2.9 &2.0\\
26b$^{25}$&4.20474&1.26&0.52&0.0 &0.0526&1.09&1.79&5590& 2.0 &+0.01&45.3 &8.6\\
&0.00005&0.13&0.05& fixed &0.0010&0.06&0.18&100&1.0 &0.13& 25.0 &1.8\\
27b$^{26}$& 3.57532 &1.007&10.39&$<$0.065&0.0476&1.05&1.08&5900&4.0&-0.1& 13.6 &4.21\\
                  &0.00006  &0.044&0.55 &                &0.0066&0.11&0.18&120  &1.0&0.1 & 4.3 &2.72\\
28b$^{27}$&5.20851&0.9550&0.484&0.0470&0.0603&1.01&1.78&5150&3.0&+0.15& $\simeq$ 30 &12.0\\
&0.00038&0.0660&0.087&0.0550&0.0050&0.14&0.11&100&0.5&0.10& &1.5\\
29b$^{27}$&2.850570&0.9&0.850&0.0820&0.0386&0.97&0.90&5260&3.5&+0.20& 4$-$13 &4.5\\
                  &0.000006&0.16&0.20&0.0810&0.0059&0.14&0.12&100&0.5&0.10& &3.5\\
30b$^{28}$&9.06005  & 1.009 & 2.90 & 0.007  & 0.0844  & 0.98 & 0.91 & 5650 & 4.3 &+0.02& 10.8 & 3.0\\
                  & 0.00024 & 0.076 &0.22&   0.031  & 0.0012  & 0.05 & 0.09 & 107  &  0.4 &0.10& 1.3 & 3.7\\
31b$^{28}$& 4.62941  & 1.46 & 0.84 & 0.02  & 0.0586  & 1.25 & 2.15 & 5730 & 2.8 & +0.00& 38.9 & 4.7\\
                  & 0.00075 & 0.30  & 0.34 &  0.16 & 0.0034  & 0.22 & 0.56 & 126  &  0.5 & 0.10& 12.7 & 2.2\\
32b$^{29}$& 6.71837  & 0.57 & 0.15 & 0.0  & 0.071  & 1.08 & 0.79 & 5970 & 3.2 & +0.00& 14.0 & - \\
                  & 0.00001 & 0.06  & 0.10 & fixed & 0.001  & 0.08 & 0.09 & 100  &  1.0 & 0.20& 6.0 & \\
33b$^{30}$&5.819143&1.10&59.0&0.0700&0.0579&0.86&0.94&5225&5.7&0.44&8.95&$>$4.6\\
&0.000018&0.53&1.8&0.0016&&0.04&0.14& 80 &0.4&0.1& & \\
\end{supertabular}
References to table:  $^{1}$\cite{Barge2008}, $^{2}$\cite{Alonso2008,Bouchy2008}, $^{3}$\cite{Deleuil2008}, $^{4}$\cite{Aigrain2008}, $^{5}$\cite{Rauer2009}, $^{6}$\cite{Fridlund2010}, $^{7}$\cite{Leger2009,Queloz2009,Barros2014,Haywood2014}, $^{8}$\cite{Borde2010}, $^{9}$\cite{Deeg2010}, $^{10}$\cite{Bonomo2010}, $^{11}$\cite{Gandolfi2010}, $^{12}$\cite{Gillon2010}, $^{13}$\cite{Cabrera2010}, $^{14}$\cite{Tingley2011}, $^{15}$\cite{Bouchy2011}, $^{16}$\cite{Ollivier2012}, $^{17}$\cite{Csizmadia2011}, $^{18}$\cite{Hebrard2011}, $^{19}$\cite{Guenther2012}, 
$^{20}$\cite{Deleuil2012}, $^{21}$\cite{Patsold2012}, $^{22}$\cite{moutou2014}, $^{23}$\cite{Rouan2012}, $^{24}$ \cite{Alonso2014}, 
 $^{25}$\cite{almenara2013}, $^{26}$\cite{almenara2013}, $^{26}$\cite{Parviainen2014}, 
$^{27}$\cite{cabrera2015},  $^{28}$\cite{Borde2018},$^{29}$\cite{Boufleur2018}, $^{30}$\cite{Csizmadia2015}\\

Despite the complexity of the CoRoT detection and vetting processes which involved different pipelines and methods that evolved during the mission lifetime and the lack of a proper estimate of the mission detection sensitivity, these planets were used to derive first-order occurrence rates \citep{Deleuil2018}. 

For close-in massive objects, brown dwarfs and hot-Jupiters, limiting the sample to those with an orbital period less than 10 days and a magnitude less than mag-r = 15.1 ensured the completeness of the follow-up process. In this range of sizes, it seems likely that only a few of them could have been missed and a detection completeness of 90\% seems reasonable. This gives  approximate occurrence rates of 0.98 $\pm$ 0.26\% for close-in hot-Jupiters and of 0.07 $\pm$ 0.05\% for brown dwarfs in the CoRoT fields. For hot-Jupiters, this result is in agreement with the occurrence rates estimated for this planet population from radial velocity surveys \citep{Wright2012, Mayor2011} but not with those from Kepler data \citep{Howard2012, Fressin2013, Santerne2016}. In particular, this is about twice the estimate derived by \cite{Santerne2016} for the Kepler giants based on planet candidate follow-up observations. On the reverse, the brown dwarf frequency estimate, treating CoRoT-3b as such, is about four times smaller than the one for the Kepler field. \cite{Santerne2016} found indeed $0.29 \pm 0.17\%$~for Kepler, but the latter covers orbital periods up to 400 days, while the three CoRoT brown dwarfs all have $P<6$\,days.

For giant planets at longer orbital period, between 10 and 100 days, the occurrence rate of 1.86 $\pm$ 0.68\% is more in agreement with studies from other sources \citep{Mayor2011, Fressin2013, Santerne2016}. Because of the typical duration of CoRoT runs and a likely much higher incompleteness of the follow-up in this range of periods, this number was calculated not only based on the confirmed planets but by including some candidates whose planetary nature could not be fully confirmed.  

The detection of CoRoT-7b \citep{Leger2009} has opened the domain of the super-Earth planets, a population that was not predicted by planet formation models but was later demonstrated by Kepler to be numerous. The super-Earth population is however at the limit of CoRoT photometric precision for solar-type stars, and CoRoT detections remained rare in this domain of small-size planets.  

While still, in principle, within CoRoT detection capability, a dearth of Neptune-sized planets was reported by \citep{Bonomo2012} from their analysis of six CoRoT fields only. This was later confirmed by \cite{Deleuil2018} who did the final analysis of all CoRoT observations. They indeed estimated that, compared to the frequency of Kepler planets for the class of small Neptunes and super-Earths \citep{Fressin2013}, the occurrence of small-size planets with Rp $<$ 5\RE\ orbiting GKM dwarfs within 10 days in CoRoT is still too low by more than a factor of three. This number was derived using the same detection completeness of 36.6 $\pm$ 6.4 \% as in \cite{Bonomo2012}. Among the various reasons, this discrepancy could have its origin in differences in the stellar population targeted by these two missions. In addition, a fraction of these small-size planets might be missed due to discontinuities in the CoRoT light curves caused by hot pixels, light curves which were still used when the final CoRoT catalog was established. With the final version of the CoRoT pipeline that corrects for a number of discontinuities being available, it will be possible to investigate this issue further.

While multi-planet systems account for about 40\% of the Kepler objects of interest (KOI), only one multi-planet transiting system, {CoRoT-24} \citep{Alonso2014}, has been reported by CoRoT. This system hosts two Neptune-sized planets. The second multi-planet system is CoRoT-7 \citep{Queloz2009}. CoRoT-7c does not transit, but its existence has been definitely established from the intensive radial velocity campaigns carried out to measure CoRoT-7b' mass \citep[e.g.]{Haywood2014}. A third planet, CoRoT-7d with a mass representative of super-Earth or Neptune size is indicated but not as well secured in the RV data \citep{Hatzes2010}. This observed low number of multi-planet detections is consistent with Kepler's results, which show that such systems are indeed numerous but are mainly found in the low mass regime of Neptune- and Earth-size planets and in the long orbital period range, a domain which is beyond the limit of CoRoT sensitivity. 

A search for circumbinary planets (CBPs, planets that orbit both components of a binary star) in CoRoT's entire sample of over 2000 EBs found three candidates based on the observation of single transits \citep{Klagyivik2017}. For the time being, none of these can be verified as a planet, however. Abundance maxima for CBPs derived from that search agree with results from the Kepler mission, which had a sample of EBs of similar size. The CBPs detected by Kepler (see Chapter "wo Suns in the Sky: The Kepler Circumbinary Planets") have periods that are mostly too long to produce multiple transits within the durations of the CoRoT runs, while their small size would have made them only marginally detectable with CoRoT. CoRoT had however good detection capability for larger or shorter-periodic (p $\lesssim$ 25 days) CBPs, but such planets either don't exist or are very rare. This survey doubled the sample size on which similar conclusions had been drawn previously \citep{Martin2015, Hamers2016}, based on searches in Kepler data.

\section{Conclusion}

The CoRoT mission has been the pioneer for space-based transiting planet detection. It was the first instrument that provided high-precision photometric observations, with almost uninterrupted coverage for weeks. Supported by an extensive program of ground-based follow-up observations, it also managed to accurately determine not only the planets diameter but also their mass and the complete set of their orbital parameters, providing observational constraints to models of formation and evolution of the close-in planet population. Surpassed by Kepler,  CoRoT accounts for a number of first results on the properties of the close-in transiting populations \citep[see][e.g.]{Moutou2013,Deleuil2018}.

The approach the CoRoT team has chosen has demonstrated the interest of combining space-based and ground-based observations in their respective domain of best performance. It also has become the {\it role model} for ESA's small satellite program and led the European community to propose both the {CHEOPS} and {PLATO} missions \citep{Broeg2013, Rauer2014}, now under development. CoRoT was designed for 2 1/2 years of operation but surpassed all expectations and continued to operate for almost 6 years producing a number of light curves similar to much more ambitious (later) missions. The legacy of CoRoT continues as is demonstrated by the number of exoplanetary systems carrying the CoRoT designation that have been further studied and continue to be studied in order to clarify all the unsolved questions relating to this topic.

\bibliographystyle{spbasicHBexo.bst}
\bibliography{HBexoCorotbib}

\end{document}